\documentclass[aps,prb,twocolumn,showpacs,superscriptaddress,groupedaddress,floatfix]{revtex4-1}
\usepackage{color}
\usepackage{mathtools}
\usepackage{latexsym}
\usepackage{graphicx}
\usepackage{float}
\usepackage{dcolumn}
\usepackage{bm}
\usepackage{amssymb}
\usepackage{amsmath}
\usepackage{mathtools}
\usepackage[space]{grffile}
\usepackage{xcolor}

\usepackage[colorinlistoftodos]{todonotes}

\newcommand{\be}{\begin{equation}}
\newcommand{\ee}{\end{equation}}
\newcommand{\bea}{\begin{eqnarray}}
\newcommand{\eea}{\end{eqnarray}}

\begin{document}


\title{Lifetime of almost strong edge-mode operators in one dimensional,
interacting, symmetry protected topological phases}

\author{Daniel J. Yates$^{1}$}
\author{Alexander G. Abanov$^{2,3}$}
\author{Aditi Mitra$^{1}$}
\affiliation{$^{1}$Center for Quantum Phenomena, Department of Physics, New York
University, 726 Broadway, New York, NY, 10003, USA\\
  $^{2}$Simons Center for Geometry and Physics, Stony Brook, NY 11794, USA\\
$^{3}$Department of Physics and Astronomy, Stony Brook University, Stony Brook,
NY 11794, USA}

\date{\today}

\begin{abstract}
  Almost strong edge-mode operators arising at the boundaries of certain
interacting 1D symmetry protected topological phases with \(Z_2\) symmetry have
infinite temperature lifetimes that are non-perturbatively long in the
integrability breaking terms, making them promising as bits for quantum
information processing. We extract the lifetime of these edge-mode operators for
small system sizes as well as in the thermodynamic limit. For the latter, a
Lanczos scheme is employed to map the operator dynamics to a one dimensional
tight-binding model of a single particle in Krylov space.
We find this model to be that of a spatially inhomogeneous Su-Schrieffer-Heeger
model with a hopping amplitude that increases away from the boundary, and a
dimerization that decreases away from the boundary. 
We associate this dimerized or staggered structure with the existence of the
almost strong mode. Thus the short time dynamics of the almost strong mode is
that of the edge-mode of the Su-Schrieffer-Heeger model, while the long time
dynamics involves decay due to tunneling out of that mode, followed by chaotic
operator spreading. We also show that competing scattering processes can lead to
interference effects that can significantly enhance the lifetime.
\end{abstract}
\maketitle

Topological states of matter are characterized by a bulk-boundary correspondence
where non-trivial topological phases host robust
edge-modes~\cite{TKNN,Bellissard94,QiRev11}. While topological phases have been
fully classified for free fermions~\cite{Ryu10}, the stability of these phases
to perturbations such as non-zero temperature, disorder, and interactions, is
poorly understood. The expectation is that as long as the perturbations are
smaller than the bulk single-particle energy gap, the edge-modes will survive.
More surprisingly, examples are beginning to emerge where even at high
temperatures of the order of the band width, and 
with moderate
interactions, the edge-modes while not completely stable, have extremely long
lifetimes~\cite{Fendley17,Nayak17,Yao19,Pollmann20}. Since
edge-modes can be used as qubits, understanding these non-perturbatively long
lifetimes is of fundamental importance both for theory and for applications.

We study a class of 1D models that in the limit of free fermions correspond to
Class-D in the Altland-Zirnbauer classification scheme~\cite{Ryu10,AZ97}. These
models host Majorana modes, and are promising candidates for non-Abelian quantum
computing~\cite{Kitaev01,Kitaev06,Fidk11,NayakRMP08,Fendley09,Alicea12,Beenaker13}.
Adding interactions and raising the temperature do not appear to destabilize the
edge-modes easily~\cite{Fendley17,Nayak17,Parker19,Yao19,Garrahan18}. Similar
behavior has been found in interacting, disorder-free, Floquet systems where
bulk quantities heat to infinite temperature rapidly, i.e., within a few drive
cycles, and yet edge-modes coexist with the high temperature bulk for an
unusually long time~\cite{Yates19}. A hurdle to understanding these lifetimes is
that they are extracted from exact-diagonalization (ED), and this is plagued by
system size effects making it difficult to extract lifetimes in the
thermodynamic limit.

We present a fundamentally new scheme to extract the long lifetimes of
topological edge-modes. Using a Lanczos scheme, we map the Heisenberg
time-evolution of the edge-mode operator onto a Krylov basis where the dynamics
is equivalent to a single-particle on a tight-binding lattice with inhomogeneous
couplings~\cite{Recbook,Altman19,Sinha19}. We find that this lattice for the
edge-mode operators is neither that of an operator of a free or integrable
model, nor is it the lattice typical of a chaotic operator. We give arguments
for the general structure of the Krylov lattice of these topological edge-modes,
and analytically extract the lifetime.

\textbf{\textit{Model:}} We study the anisotropic \(XY\) model of chain length
\(L\), perturbed by a transverse field, and by exchange interactions in the
\(z\) direction,
\begin{align}
  H&= \sum_{i=1}^L \biggl[J\left(\frac{1 + \gamma}{2}\right) \sigma_i^x
	\sigma_{i+1}^x +J\left( \frac{1 - \gamma}{2}\right) \sigma_i^y
        	\sigma_{i+1}^y\nonumber\\
        &+ J_z\sigma_i^z \sigma_{i+1}^z + g \sigma_i^z \biggr]=H_{\rm XX}+H_{\rm YY}+H_{\rm ZZ}+H_Z,
 \label{ham1}
\end{align}
where \(g\) and \(\gamma\) denote the strength of the transverse field and the
\(XY\) anisotropy respectively. We set \(J=\hbar=1\). A non-zero \(J_z\)
prevents a mapping to free Jordan-Wigner fermions.
The model has a \(Z_2\) symmetry
\(D_z = \sigma_1^z \sigma_2^z \dots \sigma_L^z\). For
\(\gamma\neq 0, J_z=0, |g| <1\), or \(\gamma \neq 0, J_z\neq 0, g=0\), the model
supports a strong mode (SM) operator defined
as~\cite{Kitaev01,Fendley12,Fendley14,Fendley16,Mila18}
\begin{equation}
    \{\Psi_0,D_z \} = 0,\qquad [H,\Psi_0] \approx u^L;\;\; ||u||<1.\label{SMdef}
\end{equation}
Thus, as \(L\rightarrow \infty\), \(\left[H,\Psi_0\right]=0\). Existence of a SM
implies that the two different parity sectors are degenerate as
\(L\rightarrow \infty\)~\cite{Kell17,Garrahan19}.

When \(J_z\neq 0\), the SM turns into an almost strong mode
(ASM)~\cite{Fendley17} that anti-commutes with parity, but only approximately
commutes with \(H\) when \(L\rightarrow \infty\). While for small system sizes,
the ASM behaves similarly to the SM as it's lifetime increases exponentially
with \(L\), for larger \(L\) however, its lifetime saturates to a system-size
independent value.

\textbf{\textit{SM for \(J_z=0\):} } While the SM is \(\Psi_0=\sigma^x_1\) when
\(J_z=g=0,\gamma=1\), for other parameters, it is a more complicated operator
which nevertheless has a finite overlap with \(\sigma^x_1\). In terms of
Majorana fermions defined as:
\begin{align}
a_{2l-1} = \sigma^x_l \prod_{j=1}^{l-1}\sigma^z_j ,\;\;
a_{2l} = \sigma^y_l \prod_{j=1}^{l-1}\sigma^z_j, \,
\end{align}
and for \(\gamma>0\),
we find that the SM localized at one end is~\cite{Suppmat}
\begin{align}
  &\Psi_0 = \sum_{l=1}^{L-1}C_l a_{2l-1};\;\; C_l = \frac{(1+\gamma)/2}{\sqrt{g^2+\gamma^2-1}}\Big(q_+^l-q_-^l\Big),\nonumber\\
  &q_{\pm}= \frac{g \pm \sqrt{g^2+\gamma^2-1}}{1+\gamma}.
  \label{SMsol}
\end{align}
$\Psi_0$ is normalizable for $g^2<1$, $\gamma \neq 0 $, indicating that it is
localized at the boundary. When \(\gamma=1\), the SM is the familiar one for the
Kitaev chain with~\cite{Kitaev01} \(C_l=g^{l-1}\). 
that, just like the correlations~\cite{franchini2005asymptotics}, the spatial
character of the SM changes at $g^2+\gamma^2=1$. 
from over-damped to under-damped decay where in the 
$g^2+\gamma^2<1$, we have 
decay of the SM does not depend 

\textbf{\textit{Autocorrelation function:}} Due to the overlap with
\(\sigma^x_1\) when \(\gamma > 0\),
the SM and ASM (together denoted by (A)SM)
might be detected through the infinite temperature auto-correlation
function~\cite{Fendley17} defined as,
\begin{equation}
    A_{\infty}(t) = \frac{1}{2^L} \text{tr} \left[\sigma_1^x(t) \sigma_1^x(0) \right].
 \label{autoc}
\end{equation}
Here \(\hat{O}(t) = e^{i H t}\hat{O}e^{-i H t}\) denotes Heisenberg
time-evolution. In general, \(A_{\infty}(t) \sim e^{-\Gamma t}\) decays in time.
For a finite wire the (A)SM can tunnel across, and the decay-rate is
exponentially dependent on \(L\) as suggested in Eq.~\eqref{SMdef}, with
\(\Gamma \sim e^{-L h(\gamma,g,J_z)}\) for some function \(h\), to be
determined. The exponential increase in lifetime with system size is a
characteristic of the SM. In contrast, for the ASM, the exponential increase of
lifetime with system size eventually saturates to a \(L\) independent result.
For example, when \(\gamma = 1\), ED suggests a highly non-perturbative
dependence \( \Gamma \sim e^{-c J/J_z}, c=O(1)\) upto logarithmic
corrections~\cite{Nayak17}. This form is also argued from setting operator
bounds on approximately conserved quantities in the prethermal
regime~\cite{Abanin17b}. However a treatment that directly studies lifetime of
topological edge-modes, and is valid for broader regimes, not necessarily
related to prethermalization, is needed.

\begin{figure}[ht]
\includegraphics[width = 0.45\textwidth]{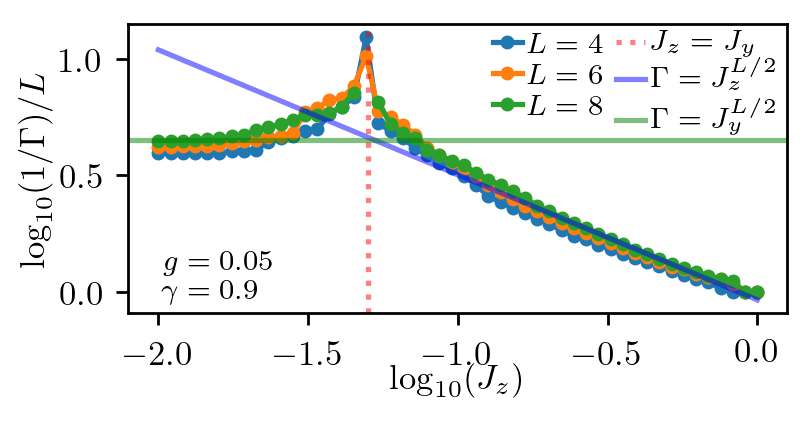}
\caption{\label{fig1} Decay-rate of \(A_{\infty}\) obtained from ED for
different \(L\) and \(J_z\), behaves as
\(\Gamma=\left({\rm max}(J_z,J_y)\right)^{L/2}\), where \(g\ll J_{z,y}\ll 1\).
When \(J_z=J_y\), destructive interference between two scattering channels leads
to 
leads to a pronounced increase of the lifetime as indicated by the cusp. }
\end{figure}

\textbf{\textit{Lifetime for small system sizes:}} We now show that the lifetime
for small system sizes is largely governed by perturbative processes. Denoting
\(|\epsilon_n\rangle\) as an eigenstate of \(H\) and parity, even in the
presence of integrability breaking terms,
\(\sigma_1^x|\epsilon_n\rangle \sim |\epsilon_n'\rangle \), where
\(\epsilon_n'\) is the opposite parity energy level nearly degenerate to
\(\epsilon_n\). Defining \(\Delta_n = \epsilon_n - \epsilon_n'\), we find that
to a good approximation the finite-size behavior is mimicked by~\cite{Suppmat}
\(A_\infty(t) \sim \sum_n \cos(\Delta_n t)/2^{L-1}\). For the finite-size
decay-rate, a perturbative estimate of \(\Delta_n\) suffices. Below we treat
\(J_{y,z},g \ll 1\), where \(J_y=(1-\gamma)/2\). Focusing on the two degenerate
ground states of \(H_{\rm XX}\) (not necessarily of definite parity) we
determine the process that gaps the states, and from that construct the two
gapped states of definite parity. The same considerations hold for every excited
level of \(H_{\rm XX}\).

Denoting the eigenstate of \(\sigma^x\) as,
\(\sigma^x|\pm\rangle = \pm |\pm\rangle\), let us first consider the case when
\(g\) is dominant. \(L\) applications of \(g\) are required for a transition
from one ground state to another,
\(|++\dots +\rangle \xrightarrow{(g\sum_i \sigma_i^z)^L} |--\dots-\rangle\).
Thus the splitting between the ground state sectors is \(g^L\), and the same
splitting appears when rotated to the basis of definite parity,
\(|++\dots +\rangle \pm |--\dots -\rangle\). The energy splitting gives a decay
rate of the ASM, \(\Gamma \sim g^{L} = e^{\log(g)L}\).

When \(J_z\) is the dominant term, the ground state degeneracy is lifted by
\(L/2\) applications of \(J_z\),
\(|++\dots +\rangle\xrightarrow{(J_z\sum_i \sigma_i^z\sigma_{i+1}^z)^{L/2}}|--\dots-\rangle\),
giving an energy splitting and consequently a decay-rate
\(\Gamma \sim (J_z)^{L/2} = e^{\log(J_z)L/2}\). Similar arguments can be applied
when \(J_y\) is dominant. Since,
\(\sigma_i^y\sigma_{i+1}^y = - \sigma_i^x \sigma_{i+1}^x \sigma_i^z \sigma_{i+1}^z\),
up to an overall sign, \(J_y\) is similar to the \(J_z\) perturbation and gives,
\(\Gamma \sim (J_y)^{L/2} =e^{\log(J_y)L/2}\). Fig.~\ref{fig1} plots \(\Gamma\)
obtained from ED. The solid lines are the estimates for \(\Gamma\) from
perturbation theory, and they excellently describe the asymptotic behavior of
the data. In addition, the plot shows an interesting 
phenomenon when competing terms affect the lifetime. In particular, when
\(J_y \sim J_z\), since matrix elements of the two terms have opposite signs,
destructive interference between these two scattering channels leads to an
enhanced lifetime. This is visible as a pronounced cusp in Fig.~\ref{fig1} when
\(J_y\sim J_z\).

\textbf{\textit{Krylov basis:}} We now discuss the lifetime of the ASM in the
system-size independent limit. We study the operator dynamics following a
Lanczos scheme designed to map the Heisenberg time-evolution to a tight-binding
model in Krylov space~\cite{Recbook,Altman19}. Note that,
\(\sigma_1^x(t) = e^{i H t} \sigma_1^x e^{-i H t} = \sum_{n = 0}^\infty \frac{\left(i t \right)^n}{n!}\mathcal{L}^n \sigma_1^x\),
where \(\mathcal{L} = \left[H,\cdot\right]\). We define \(\hat{O} = |O)\), and
\((O_1|O_2) = \frac{1}{2^L}\text{tr}\left[O_1^\dagger O_2 \right]\). Thus
\(A_\infty(t)\) becomes,
\begin{equation}
  \label{ainf_L} A_\infty(t) = (\sigma_1^x| e^{i t\mathcal{L}} |\sigma_1^x).
\end{equation}
The Pauli basis is \(2^{2L}\) dimensional, hence determining
\(\mathcal{L}\) outright is generally not feasible. \(\sigma_1^x\) is a
Majorana, and when the system is free, the time evolution can only mix with a
total of \(2L\) Majoranas, and only in this case can \(\mathcal{L}\) be
completely determined. However, the key observation is that for both free and
interacting cases, there is a special basis, the Krylov basis, where
\(\mathcal{L}\) is tri-diagonal, and the dynamics of any operator can be mapped
to a tight-binding model.

To construct the Krylov basis for say \(\sigma^x_1\), we start with
\(|O_0) = |\sigma_1^x)\), and construct \(|A_1) = \mathcal{L}|O_0) \),
\(b_1 = \sqrt{(A_1|A_1)}\), and \(|O_1) = |A_1)/b_1\). These steps are repeated
as follows,
\begin{align}
  &|A_n) = \mathcal{L} |O_{n-1}) - b_{n-1} |O_{n-2}),\nonumber\\
  &b_n = \sqrt{(A_n|A_n)}; |O_n) = \frac{1}{b_n}|A_n).
\end{align}
In the Krylov basis, the Liouvillian takes the form,
\begin{equation}
  \mathcal{L} =H_K= \sum_i b_i\left(c_i^{\dagger}c_{i+1}+ h.c.\right),
 \label{HK}
\end{equation}
where \(c_i^{\dagger},c_i\) are the creation, annihilation operators in the 
Krylov basis. Recently, this approach has been mainly used to identify
chaos~\cite{Altman19,Sinha19,Gorsky19}. Below we show that this method is very
helpful for studying long-lived topological edge-modes.

\textbf{\textit{Lifetime in the thermodynamic limit:}} The (A)SM can be
constructed by noticing that
\begin{align}
  &[H_K,c_1^\dagger] = b_1 c_2^\dagger,\;\;\; [H_K,c_1^\dagger - \frac{b_1}{b_2}c_3^\dagger]
    = -\frac{b_1b_3}{b_2}c_4^\dagger,\nonumber\\
  &[H_K,c_1^\dagger - \frac{b_1}{b_2}c_3^\dagger + \frac{b_1 b_3}{b_2 b_4} c_5^\dagger]
    = \frac{b_1b_3 b_5}{b_2 b_4}c_6^\dagger\dots.
\end{align}
Thus the ASM after \(N\) iterations is,
\begin{align}
  \label{my_bns}
  \Psi_0(N)= \sum_{n = 0}^{N}
  (-1)^n \frac{b_1 b_3 \dots b_{2n-1}}{b_2 b_4 \dots b_{2n}} c_{2n+1}^\dagger.
\end{align}
The error, defined by how much \(\Psi_0(N)\) does not commute with \(H_K\) is
\begin{equation}
  \!\! \text{error}(N) = [H,\Psi_0(N)]\!\! =
  (-1)^N \frac{b_1 \dots b_{2N-1}b_{2N+1}}{b_{2}\dots b_{2N}}c_{2N+2}^{\dagger}.
 \label{eqn_error}
\end{equation}
The error is an important quantity for identifying an (A)SM. This is because for
a SM, the error only decreases with subsequent iterations, whereas for an ASM,
the error decreases up to a certain \(N^*\), and then begins to grow. In
addition, as we show below, the error at \(N^*\) can be used to determine the
lifetime in the thermodynamic limit.
\begin{figure}[ht] \includegraphics[width = 0.45\textwidth]{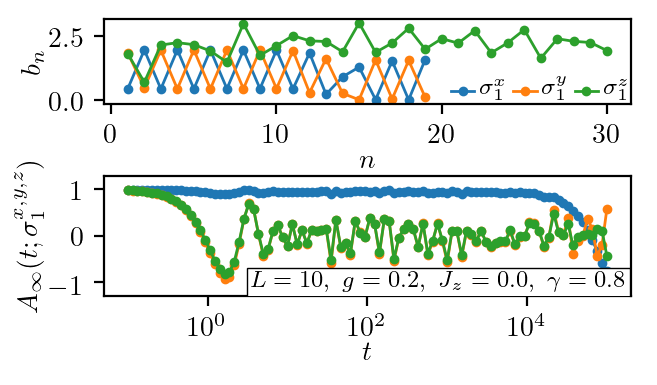}
\caption{\label{fig2} Top panel: The \(b_n\) for the Pauli spins on the first
site \(\sigma_1^{x,y,z}\) for \(J_z=0,\gamma>0\). The model maps to free
fermions and supports a SM with overlap with \(\sigma^x_1\). The deviation from
perfect staggered behavior for \(n>10\) is a finite-size effect. Bottom panel:
\(A_{\infty}\) from ED for \(\sigma^{x,y,z}_1\). Due to overlap with the SM,
\(\sigma^x_1\) persists up to \(t\sim 10^4\) as opposed to \(\sigma^{y,z}_1\)
that decay by \(t=O(1)\). }
\end{figure}

First consider \(J_z=0, g^2<1\), for which a SM exists (c.f. Eq.~\eqref{SMsol}).
We find that the Krylov Hamiltonian for \(\sigma^x_1\) with \(\gamma=1\) is,
\(b_{\rm odd}= 2g, b_{\rm even}= 2\), and therefore has a staggered/dimerized
structure quantified by \(b_{2n}-b_{2n+1}> 0\). For \(\gamma\neq 1\), the
\(b_n\) are shown in Fig.~\ref{fig2}, and show a similar staggered structure.
Thus the effective Hamiltonian in the Krylov basis is the Su-Schrieffer-Heeger
(SSH) model~\cite{SSH79,SSH80}, with the SM being the edge-mode of the SSH
model. For the same parameters, other Pauli operators such as \(\sigma^{y,z}_1\)
which are not localized at the edge under Heisenberg time-evolution, have a
qualitatively different Krylov Hamiltonian. In particular, \(\sigma^y_1\) is
given by an SSH-type model but with a dimerization of the opposite sign to that
of \(\sigma^x_1\), so that the effective Hamiltonian for \(\sigma^y_1\) is
topologically trivial and supports no localized edge-mode. Since topological
protection is robust to moderate disorder, local fluctuations of the above
staggered structure in Krylov space will not affect the stability of the
edge-mode. The pattern of staggering of $b_n$ in Fig.~\ref{fig2} continues until
$n\sim O(L)$, after which finite-size effects, such as the hybridization of the
Majoranas at the ends of the chain, set in.

The Krylov basis for \(\sigma^z_1\) is different from \(\sigma^{x,y}_1\) in that
to start with, near site \(1\) the dimerization is negative, corresponding to a
topologically trivial phase. But on moving towards the bulk, the average hopping
first increases, and then plateaus. The net effect on the dynamics is similar to
that on \(\sigma^y_1\) in that this lattice causes the operator to spread
rapidly into the bulk under time-evolution. The lower panel of Fig.~\ref{fig2}
shows the \(A_{\infty}\) of the 3 Pauli operators, with \(\sigma^{y,z}_1\)
decaying rapidly.

\begin{figure}[ht]
\includegraphics[width = 0.45\textwidth]{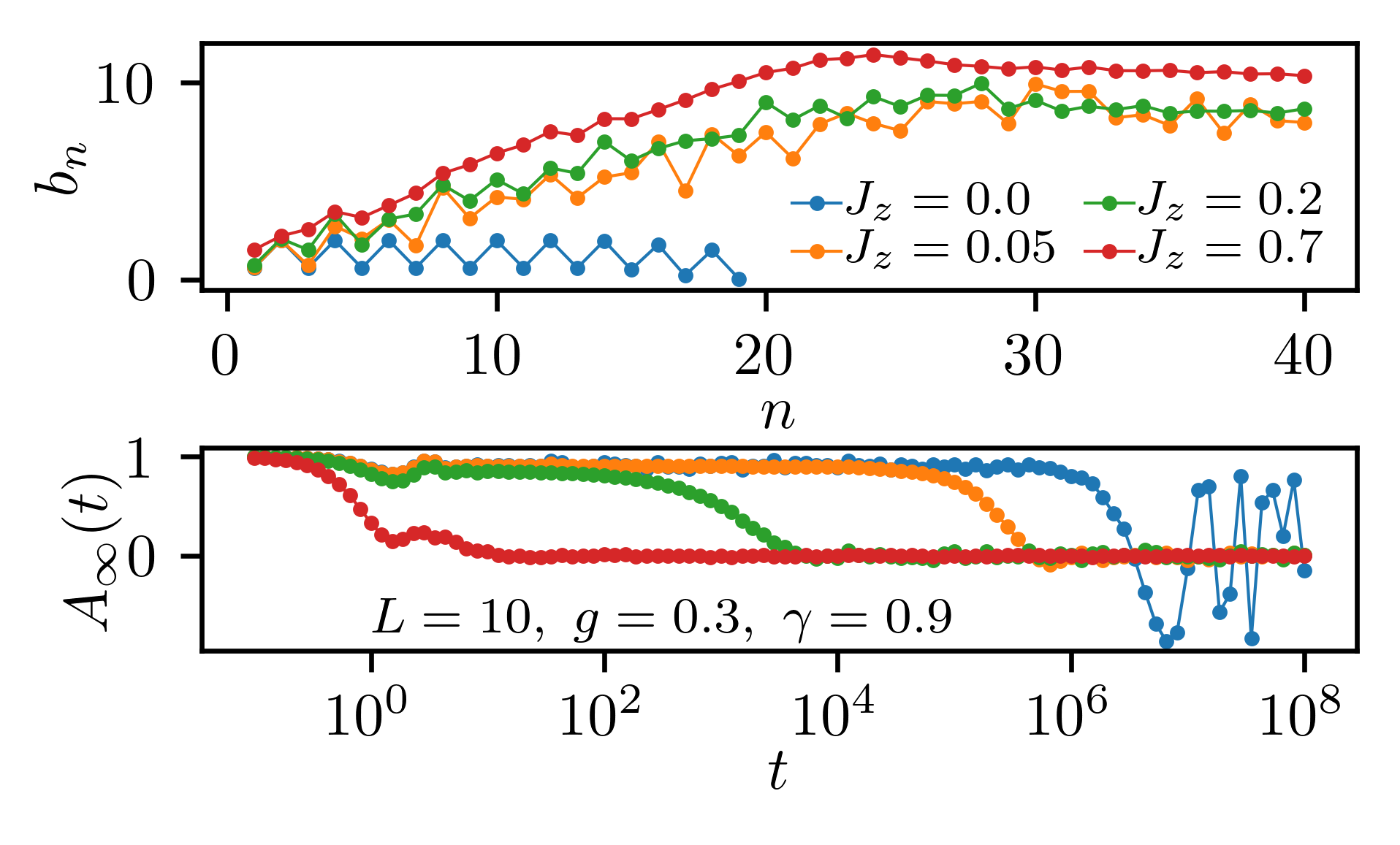}
\caption{\label{fig3}
Top panel: The \(b_n\) for increasing \(J_z\) with finite-size effects appearing
as a plateau for \(n>20\). As \(L\) is increased~\cite{Suppmat}, the linear ramp
is extended, with the \(b_n\) plateauing at a larger \(n\) and at a larger
value. Bottom panel: \(A_{\infty}\) shows rapid decrease in lifetime with
increasing \(J_z\). }
\end{figure}

 Fig.~\ref{fig3}, top panel shows how the \(b_n\) change on increasing \(J_z\).
The corresponding \(A_{\infty}\) is plotted in the lower panel of
Fig.~\ref{fig3}. One finds that the effect of \(J_z\) is two-fold, one is to
increase the average hopping into the bulk, which appears as a non-zero slope of
\(b_n\) when plotted against \(n\). The second effect is to reduce the
dimerization with increasing \(n\). Eventually, deep in the bulk, the
dimerization vanishes, and the effective hopping increases linearly with
position, a behavior expected for a generic chaotic
operator~\cite{Altman19,Sinha19}. The long lifetime of the ASM is entirely due
to this crossover from the topologically non-trivial SSH model at small \(n\),
to chaotic linear couplings at large \(n\).

\textbf{\textit{Effective model in Krylov basis:}} We make this more
quantitative by adopting the following model for the hopping parameters
\begin{align}
    b_{2n} &= \alpha_0 + \rho \alpha 2n + \delta; \quad &0<\alpha\ll \alpha_0 \sim \delta;
 \nonumber \\
    b_{2n+1} &= \alpha_0 + \alpha (2n+1); \quad &0<1-\rho \ll 1\,.
 \label{2slope}
\end{align}
The even sites have slope \(\rho \alpha\), while the odd sites have slope
\(\alpha\). At some point the dimerization \(b_{2n}-b_{2n+1}\) changes sign as
\(\rho<1\). This means that Eq.~\eqref{eqn_error} eventually grows with $N$ and
the mode is non-normalizable. We also imposed \(1-\rho<<1\) to simplify
analytic expressions but this restriction is not essential.

We can estimate the decay-rate from Eq.~\eqref{eqn_error} by finding \(N^*\)
such that \(b_{2N^*+1}= b_{2N^*}\) which gives,
\(N^*\sim \delta/2 \alpha (1-\rho )\gg 1\) and,
\begin{align}
  &\Gamma \sim |\text{error}(N^*)|=
    b_1 \exp\left[\sum_{n=1}^{N^*}\ln \biggl(\frac{b_{2n+1}}{b_{2n}}\biggr)\right] \nonumber\\
  &\sim \exp\left[ -\frac{\delta}{2 \alpha}\log\left(\frac{1 }{1-\rho}\right) \right]
    \label{LT1}.
\end{align}
Note that when \(\rho=1,\alpha\neq 0\), we still have a SM despite the fact that
the \(b_n\) have a linear slope \(b_n \sim \alpha n\). Thus it is the
dimerization, which is preserved when \(\rho=1\), that prevents the operator
from spreading. Eq.~\eqref{LT1} shows that the lifetime depends on \(J_z\)
non-perturbatively as the slope \(\alpha \propto J_z\). We later give numerical
and qualitative arguments for this form of the slope.

It is illuminating to consider the continuum limit of the effective Hamiltonian
in the Krylov basis, where the the eigenvalue problem may be recast
as~\cite{Suppmat},
\( E \Psi_n = \biggl[\left(b_{2n-1} - b_{2n}-b_{2n}\partial_n\right)\sigma^- + h.c \biggr]\Psi_n\).
The edge-mode solution is,
\begin{align}
    \Psi_{0,n} = e^{-\int ^{n}_{1} dm\frac{b_{2m} - b_{2m-1}}{b_{2m}}}
    \begin{pmatrix}1 \\ 0 \end{pmatrix},
 \label{ASM1}
\end{align}
and shows that the ASM, indeed, decreases in amplitude into the bulk when
\((b_{2m}-b_{2m-1})/b_{2m}>0\). Using the minimal model in Eq.~\eqref{2slope},
we see that at \(N^*\), \(b_{2N^*}-b_{2N^*+1}=0\), ~\eqref{ASM1} stops
decreasing with $n$ and mixes with other modes. The decay-rate is estimated by
the value of ASM at $n=N^*$,
\(\Gamma \sim \exp\left[-\int^{N^*} dm \frac{b_{2m}-b_{2m-1}}{b_{2m}}\right]\)
which recovers Eq.~\eqref{LT1}.

\begin{figure}[ht]
\includegraphics[width = 0.45\textwidth]{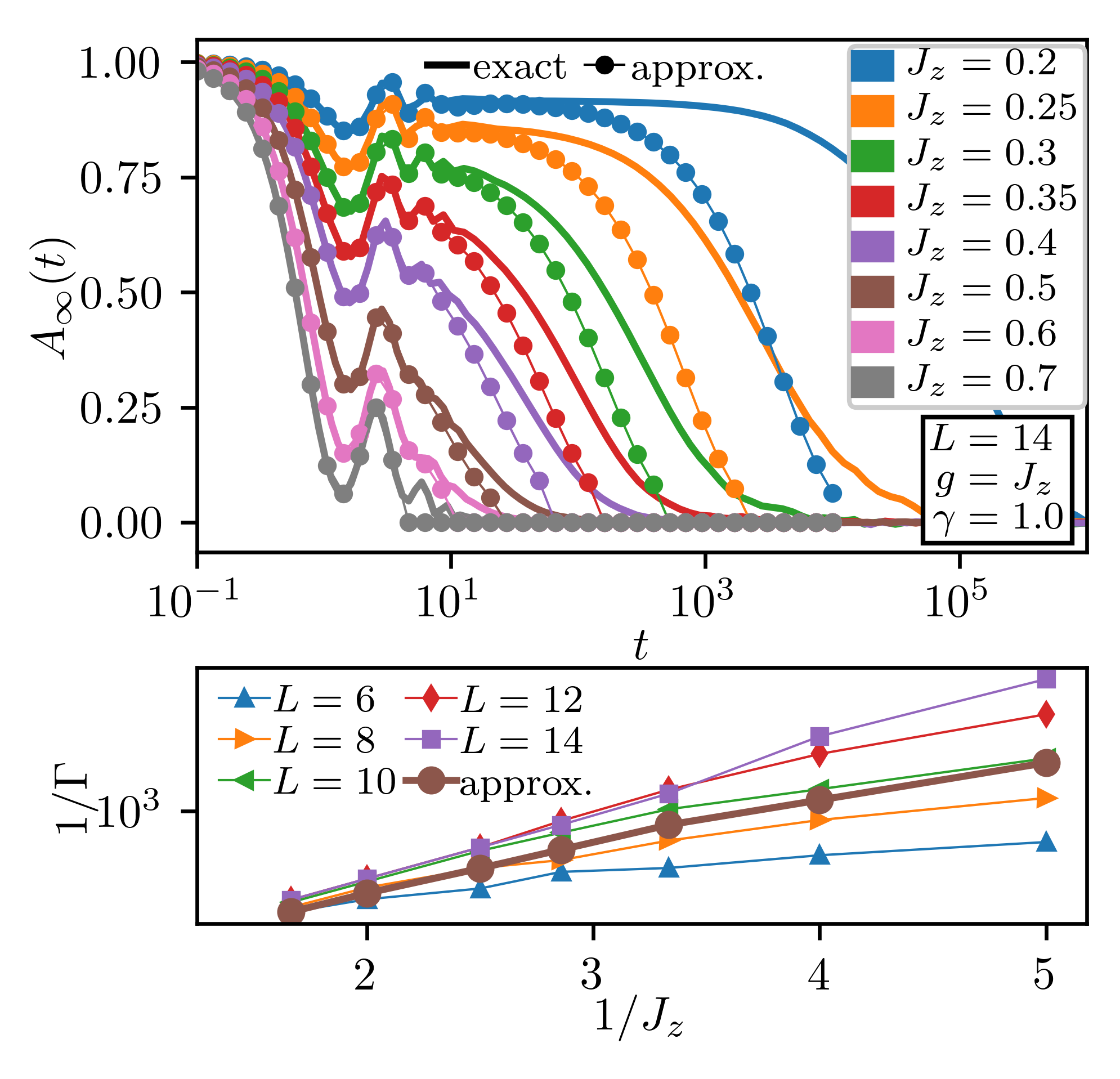}
\caption{\label{fig4}
 Top panel: \(A_{\infty}\) from ED for \(L=14\), with \(J_z=g\) increasing from
the top to the bottom of the panel. It is compared with the approximate
\(A_{\infty}\) obtained from time-evolution in a Krylov basis with \(\sim 40\)
exact \(b_n\)~\cite{Suppmat}, and \(b_{40<n<n_{\rm max}}=b_{40}\) held constant,
mimicking a reservoir. We choose \(n_{\rm max} = 2e5\), which is large enough to
capture the decay of the ASM before finite-size effects set in. Bottom panel:
Decay-rate extracted from both numerical methods. For \(1/J_z<4\), the overlapping
lines for \(L=12,14\) indicates saturation of the lifetime.}
\end{figure}

\textbf{\textit{ Comparison of ED with Krylov Hamiltonian with a metallic
bulk:}} We extract the non-perturbative lifetime using two different numerical
methods. Top panel of Fig.~\ref{fig4} compares \(A_{\infty}\) from ED for
\(L=14\), to that obtained from time-evolving by the Krylov Hamiltonian
\(\langle n=1|\left[\exp\left(i H_K t\right)\right]|n=1\rangle\), where
\(|n=1\rangle\) is a state localized at site 1 in the Krylov basis. Since the
calculation of the \(b_n\) is exponentially expensive in computer resources,
only the first \(\sim 40\) \(b_n\) are evaluated.
Guided by Fig.~\ref{fig3}, we simulate a semi-infinite lattice in Krylov space
by setting \(b_{40<n<2e5} = b_{40}\), essentially attaching a metallic reservoir
to our inhomogeneous SSH model. The lifetime obtained by both these methods is
shown in the lower panel of Fig.~\ref{fig4}, and suggests the \(L\) independent
form \(\ln\Gamma \propto -1/J_z\). Thus for the purpose of capturing the
lifetime, the simple model for the bulk \(b_n\) is an efficient alternative to
ED. In addition, the saturation of the lifetime implies that it is controlled by
the dimerization of the \(b_n\) at small and intermediate \(n\). \cite{Suppmat}

\textbf{\textit{Qualitative argument for \(\alpha \propto J_z\):}} We supplement
the above results for the decay-rate by a qualitative argument for
\(\alpha \propto J_z\). For simplicity we restore \(J\) and consider
\(\gamma=1\). When \(J_z=g=0\), then \(H=H_{\rm XX}=J N\) counts the number of
domain walls \(N=\sum_i\sigma^x_i\sigma^x_{i+1}\). When \(J_z\neq 0\) we recast
\(H_{ZZ} = J_z D + J_z \tilde{E}\), where \(D\) commutes with \(N\), whereas
\(\tilde{E}\) does not. We find that the operator,
\begin{align}
  D = \sum_j P_j \sigma^z_j \sigma^z_{j+1};\;\; P_j= \frac{1}{2}\biggl[1-\sigma^x_{j-1}\sigma^x_{j}\sigma^x_{j+1}\sigma^x_{j+2}\biggr],
\end{align}
does not change the number of domain walls and commutes with $N$\cite{Suppmat}.
\(D\) is essentially a hopping term for domain walls. In the basis that
simultaneously diagonalizes \(D,N\) we find that the minimal energy to create a
domain wall in the bulk is reduced from \(2J\) to \(2J-J_z\), and that domain
wall particle-hole pairs have energies of O($J_z$). Now consider \(J_z\ll g\).
Then the leading term non-commuting with \(N\) is \(H_Z\).
 
As argued for a different model~\cite{Yao19}, the energy cost for flipping a
spin at the edge is \(\sim J\). Thus a creation of \(\sim J/J_z\) pairs of
domain walls in the bulk can off-set the energy \(J\) required to flip an edge
spin. This requires \(J/J_z\) applications of the transverse field $g$.
Therefore the Fermi-Golden rule estimate for the decay rate is,
\begin{align}
  \Gamma \sim g \biggl[\frac{g}{J}\biggr]^{cJ/J_z}, \;\;\;c=O(1).
 \label{gres}
\end{align}
Upto logarithms, this decay-rate is consistent with ED~\cite{Nayak17}
(Fig.~\ref{fig4}), operator bounds in the prethermal regime~\cite{Abanin17b},
and with time-evolution using a truncated Krylov Hamiltonian (Fig.~\ref{fig4}).

\textbf{\textit{Summary:}}
We have presented a new way to extract the non-perturbatively long lifetimes of
ASMs. We showed that the Krylov Hamiltonian for the ASM has linearly growing
hopping along with decreasing dimerization, where the dimerization is associated
with the existence of the ASM and is key to preventing chaotic operator growth.
Essentially the operator dynamics is that of a particle which is trapped for a
long time as a quasi-stable SSH edge-mode that eventually escapes via tunneling.
We demonstrated that a truncated Krylov Hamiltonian terminating in a metallic
bulk is an efficient way for capturing the lifetime of the ASM. We also found
that competing terms can interfere to enhance the lifetime (Fig.~\ref{fig1}). It
would be interesting to identify additional structures of the Krylov
Hamiltonian, besides dimerization, that can support long-lived edge-modes. More
broadly, generalization of this study to other topological states, both static
and Floquet, and in any spatial dimension, is an exciting avenue for future
research.

{\sl Acknowledgements:}
This work was supported by the US Department of Energy, Office of
Science, Basic Energy Sciences, under Award No.~DE-SC0010821 (DJY and AM)
and by the US National Science Foundation Grant NSF DMR-1606591 (AGA).


%

\begin{widetext}
  \section*{Supplementary Material: Lifetime of almost strong edge mode operators in one dimensional, interacting,
    symmetry protected topological phases}

  The supplementary material contains:\newline
1. Three supplementary plots.\newline
2. Derivation of SM for general \(\gamma\).\newline
3. Derivation of the continuum model. \newline
4. Derivation of Eq.~(16).

\section*{Figures showing the autocorrelation function of \(\sigma^x_1\) and also the effective Krylov hoppings}

In this section we present three additional plots for the autocorrelation function, and for the $b_n$ parameters of the Krylov Hamiltonian.

The existence of the (almost) strong mode (A)SM leads to the near degeneracy of energy levels of opposite parity. Let us denote \(|\epsilon_n\rangle\) as an eigenstate of \(H\) and parity. Then, even in the presence of integrability breaking terms, \(\sigma_1^x|\epsilon_n\rangle \sim |\epsilon_n'\rangle \), where
\(\epsilon_n'\) is the opposite parity energy level nearly degenerate to
\(\epsilon_n\). Defining \(\Delta_n = \epsilon_n - \epsilon_n'\), we find
that to a good approximation the finite-size behavior is mimicked by
\begin{align}
    A_\infty(t) \sim \sum_n \cos(\Delta_n t)/2^{L-1}\,.
 \label{neardeg}
\end{align}

Fig.~\ref{fig5} shows the exact autocorrelation function obtained from ED and compares it with the approximation \eqref{neardeg} computed for 
$\Delta_n = \epsilon_n - \epsilon_n'$ where the level $\epsilon_n' \equiv \epsilon_m $ is found with 
the following relation,
$\text{argmax}_{m}|\langle m | \sigma_1^x |n\rangle|^2 $. 
One can see that this approximation reproduces the lifetime not only for small system sizes, but also for larger systems where the lifetime has saturated (ie,
become \(L\) independent). 
\begin{figure}[ht]
\includegraphics[width =3.4in]{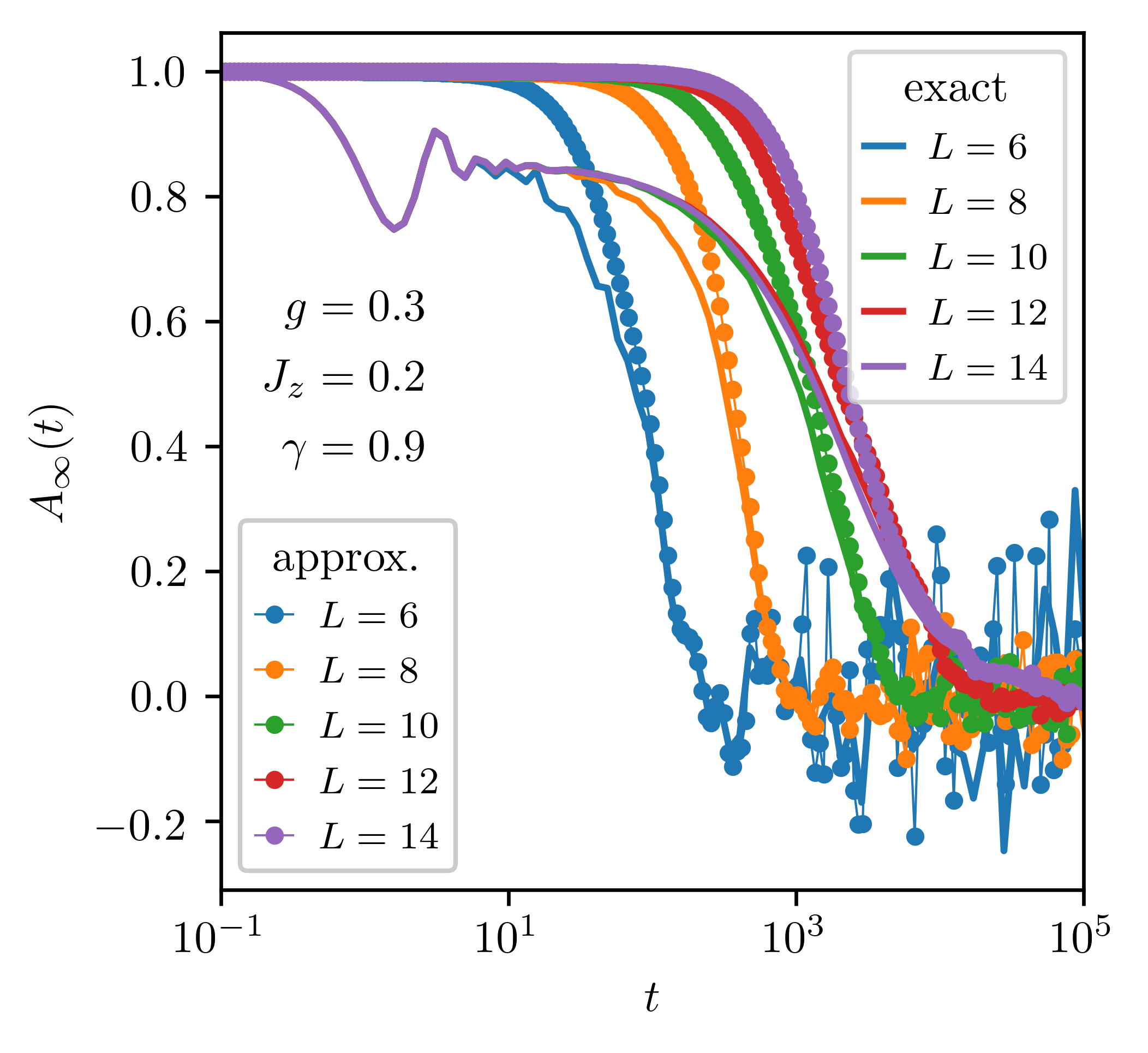}
\caption{\label{fig5} 
 Exact autocorrelation function obtained from ED and compared with the approximation \(A_\infty(t) \sim A_\infty^C(t)=\sum_n \cos(\Delta_n t)/2^{L-1}\).
 If for nearly all eigenstates \(|n\rangle \), there exists another eigenstate state \(|m(n)\rangle \), 
 such that, \(|\langle m(n) |\sigma_1^x |n\rangle |^2 \sim 1\), then \(A^{C}_\infty\) is a good approximation, and 
 estimates the lifetime well. In fact it reproduces the lifetime before system  size saturation, as well as the system size independent results.
}
\end{figure}

Fig.~\ref{fig6} shows the \(b_n\) for different \(J_z\) and for different system sizes. 
It also shows the corresponding \(A_\infty(t)\) from ED for the same parameters. The figure suggests that for chains exhibiting an anomalously long lifetime in the autocorrelation function, the Krylov parameters \(b_n\) have three main features, a ramp upwards at small \(n\), 
a system-size dependent plateau at intermediate and large \(n\), and 
\(J_z, g, \gamma,n\) dependent staggering  or dimerization of the \(b_n\).
\begin{figure}[ht]
\includegraphics[width =7.2in]{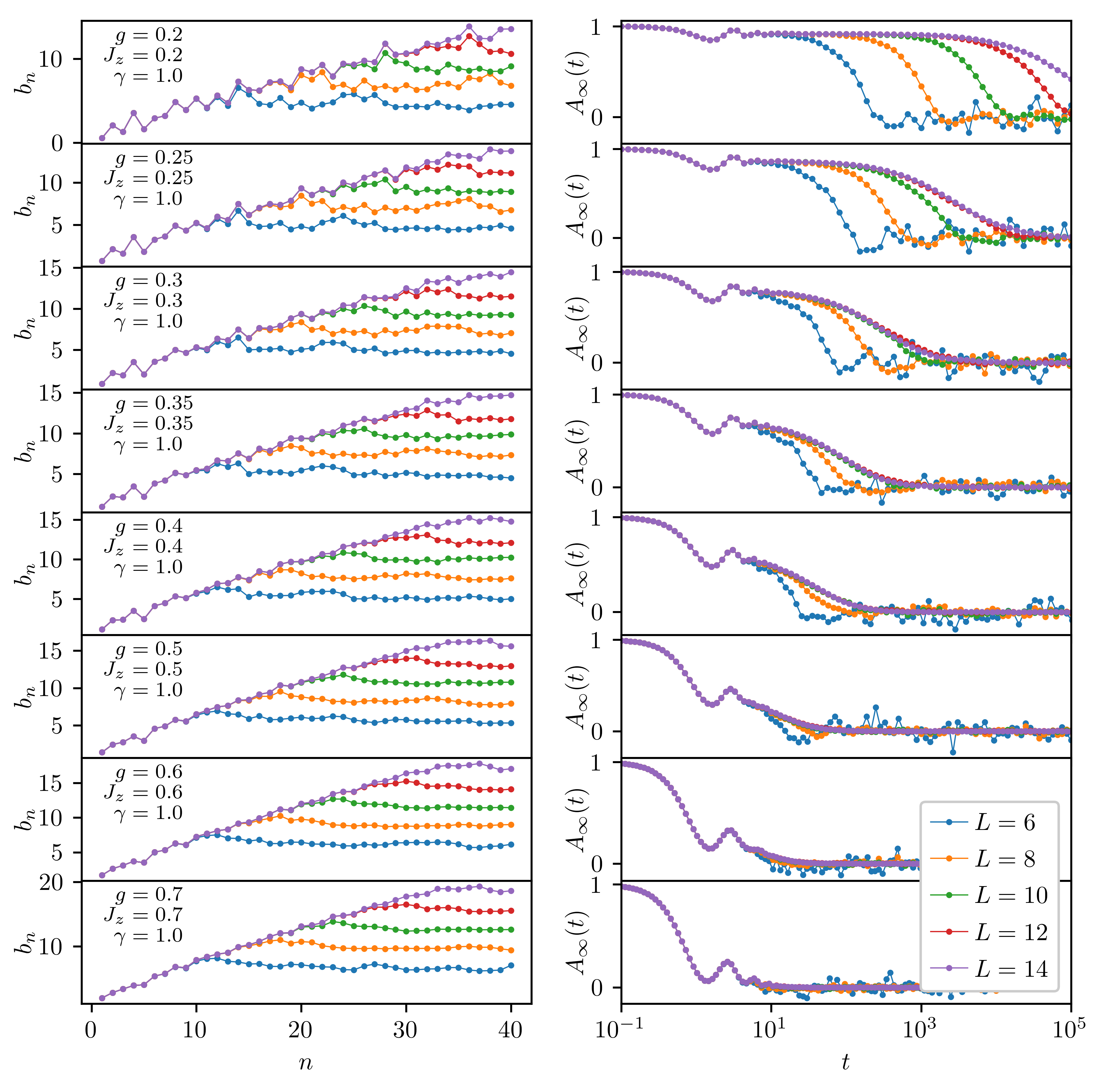}
\caption{\label{fig6} 
Left column, \(b_n\) for different \(J_z\) and for different system sizes. 
We also take \(g=J_z\) as in Fig.~4 in the main text.
Right column, \(A_\infty(t)\) from ED for the same parameters as the left column.
All \(A_\infty(t)\) show saturation in system size for \(J_z\geq 0.25\). 
The \(b_n\) have three main features, a ramp upwards at small \(n\), 
a system-size dependent plateau at intermediate and large \(n\), and 
\(J_z, g, \gamma,n\) dependent staggering or dimerization of the \(b_n\).
The Krylov subspace of \(\sigma_1^x\) is \( \propto 2^{2L} \) and 
generally, the plateaus in the the left column extend out to 
very large \(n\). As the \(b_n\) are exponentially difficult to 
compute, only the first 40 are shown. Knowledge of all \(b_n\) and 
usage of Eq.~(6) will reconstruct the right column exactly.
From top to bottom, both sides, \(J_z\) is increased. The increase 
in \(J_z\) drastically reduces the lifetime of \(A_\infty(t)\) while
simultaneously diminishing the staggering of the \(b_n\). In particular, 
the onset of a ``smooth'' \(b_n\) structure is moved to smaller \(n\) as 
one increases \(J_z\).
}
\end{figure}

Fig.~\ref{fig7} shows how \(b_n\) varies for different \(J_z\), for system size \(L=14\). The overall staggering in \(b_n\) is reduced as one increases \(J_z\), and 
the staggering is also stronger at smaller \(n\). We associate this staggered structure at small \(n\) with the existence of the ASM.  

\begin{figure}[ht]
\includegraphics[width =3.4in]{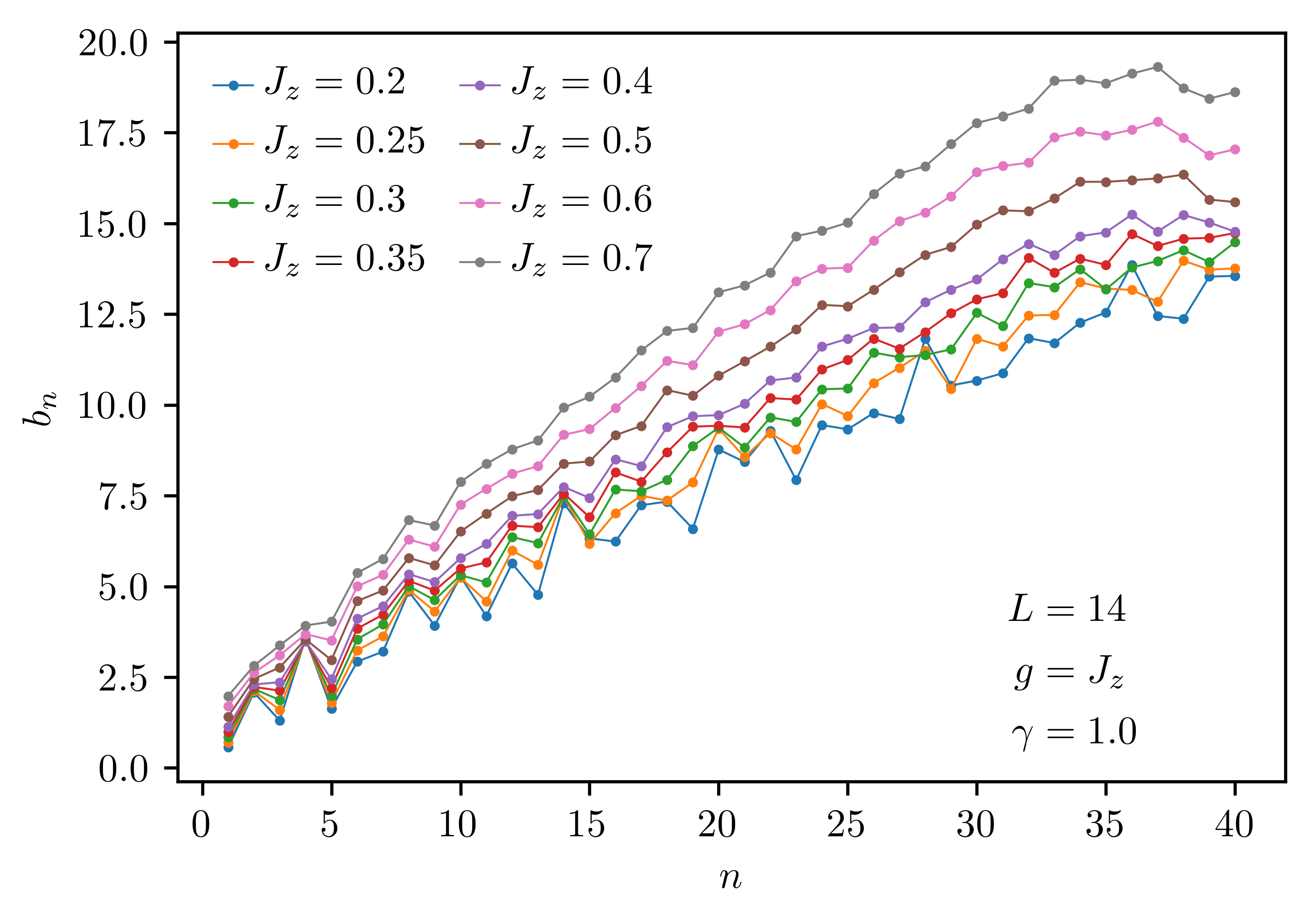}
\caption{\label{fig7} 
The \(b_n\) for different \(J_z\) for system size \(L=14\) shown in Fig.~\ref{fig6}.
Overall staggering in \(b_n\) is reduced as one increases \(J_z\), and 
the staggering is also stronger at smaller \(n\).
The approximate \(A_{\infty}\) shown in Fig.~4 in main text is constructed 
from these \(b_n\) followed by an approximate ``plateau'' of \(b_{n>40} = b_{40}\) 
for \(n\) from 41 to 200,041. 
}
\end{figure}

\section*{Constructing zero-mode for the \(XY\) model \(J_z=0\)}

For $J_z=0$ the model (1) in the main text can be reduced to a model of non-interacting Majorana fermions.  Defining, 
\begin{equation}
    a_{2l-1} =  \sigma^x_l \prod_{j=1}^{l-1}\sigma^z_j,\qquad
    a_{2l} =  \sigma^y_l \prod_{j=1}^{l-1}\sigma^z_j,
\end{equation}
we obtain from (1) in main text
\begin{align}
    H = i\sum_{l=1}^L\left[-\frac{1+\gamma}{2}a_{2l}a_{2l+1}+\frac{1-\gamma}{2}a_{2l-1}a_{2l+2}-g a_{2l-1}a_{2l}\right] \,.
\end{align}
Here one should assume $a_{2L+1}=a_{2L+2}=0$, as this ensures that
\(\sigma^{x,y}_{L+1}\) is outside the system.
It is straightforward to construct the operators $\Psi_k$ such that \(\left[H,\Psi_k\right]=E_k\Psi_k\), with \(\Psi_k|0\rangle\)
creating the eigenstate \(|k\rangle\) from vacuum. 
The spectrum is given by 
\begin{align}
    E_k=\pm 2\sqrt{(g+\cos k)^2+\gamma^2\sin^2k}\,,
 \label{bulk-spectrum}
\end{align} 
and has a gap for $\gamma\neq 0$ with eigenstates given by superpositions of right and left propagating waves of wave-vector \(k\). The bulk spectrum \eqref{bulk-spectrum} is essentially the one for spin chains with periodic boundary conditions. However, for open boundary conditions, there is also the possibility of having bound mid-gap eigenstates. Let us look for the states with zero energy $E=0$ corresponding to the following operators\footnote{Strictly speaking the energy is exactly zero only for a half-infinite chain. For a long but finite chain the energy is exponentially small in the length of the chain}: 
\begin{equation}
  \Psi_0^{+} =\sum_{l =1}^{L-1}C_{l}^{+} a_{2l-1},
  \qquad
  \Psi_0^{-} =\sum_{l =1}^{L-1}C_{l}^{-} a_{2l}\,.
\end{equation}
Then requiring \(\left[H,\Psi_0^{\pm}\right]=0\) gives,
\begin{align}
  g C_l^\pm - \left(\frac{1 \pm \gamma}{2}\right) C_{l+1}^\pm 
  -\left( \frac{1 \mp \gamma}{2}\right)C_{l-1}^\pm=0.\label{rec1}
\end{align}
Note that the two recursion relations are mapped to one-another via the 
inversion symmetry operator, \(i \rightarrow L -i \), thus we expect,
\(\Psi_0^\pm\) to yield edge modes on the left and right ends of the 
wire.
Imposing that \(C_l^+ \propto u^l, C_l^- \propto v^l\) yields solutions 
\(u_\pm, v_\pm\),
\begin{align}
  u_{\pm} &= \frac{1}{1 + \gamma} \biggl[g \pm \sqrt{g^2+\gamma^2 -1}\biggr],\\
  v_{\pm} &= \frac{1}{1 - \gamma} \biggl[g \pm \sqrt{g^2+\gamma^2 -1}\biggr].
 \label{abpm}
\end{align}

We now construct the edge mode on the left end of the wire by imposing the boundary condition \(C_0^{\pm}= 0\) and  fixing \(C_1^{\pm} = 1\), 
\begin{align}
    C_l^+ &= \frac{1 + \gamma}{2 \sqrt{g^2 + \gamma^2 -1}} \left(u_+^l - u_-^l \right),\\
    C_l^- &= \frac{1 - \gamma}{2 \sqrt{g^2 + \gamma^2 -1}} \left(v_+^l - v_-^l \right).
 \label{Clpm}
\end{align}

When \(g^2>1\), \(C_l^\pm\) yields a growing solution, regardless of \(\gamma\). 
Thus the solution is a
non-normalizable operator as \(L \rightarrow \infty\), and no zero mode exists.
When \(g^2<1, \gamma>0\), \(C_l^+\) yields a normalizable solution, \(C_l^-\) does not. On imposing appropriate boundary conditions \(C_l^-\) will give the zero mode on the right end of the chain. 
When \(g^2<1, \gamma<0\), \(C_l^-\) yields a normalizable solution, \(C_l^+\) does not (or rather \(C_l^+\) is related to the zero mode at the right end of the chain).

With the above observations, we define \(q_\pm\),
\begin{equation}
    q_\pm = \frac{1}{1 + |\gamma|} \biggl[ g \pm \sqrt{g^2 + \gamma^2 -1} \biggr]=\theta(\gamma) u_{\pm} + \theta(-\gamma)v_{\pm},
\label{Qlpm}
\end{equation}
and we drop the \(\pm\) label on \(C_l^\pm\)
\begin{equation}
    C_l = \frac{1 + |\gamma|}{2 \sqrt{g^2 + \gamma^2 -1}} \left(q_+^l - q_-^l \right).
    \label{newcl}
\end{equation}
Our edge operator on the left end becomes,
\begin{equation}
\Psi_0 = \sum_{l = 1}^{L-1} C_l \biggl[ \theta(\gamma) a_{2l-1} + \theta(-\gamma) a_{2l} \biggr],
\end{equation}
reproducing (4) in the main text.
In summary, for \(g^2>1\), we are in a trivial phase. For \(g^2<1,\gamma>0\),
we have a zero mode with overlap with \(\sigma^x_1\). For \(g^2<1, \gamma<0\), we have a zero mode which now overlaps with \(\sigma^y_1\) rather than \(\sigma^x_1\).

It is clear from (\ref{Qlpm},\ref{newcl}) that the spatial character of the edge modes change at $g^2+\gamma^2=1$ from  over-damped  to  under-damped decay.  In the under-damped regime $g^2+\gamma^2<1$, we have $|q_\pm|^2=\frac{1\mp \gamma}{1\pm \gamma}$ and the amplitude (\ref{newcl}) in position space oscillates and decays/grows with a rate which is independent of $g$.
 Not surprisingly, overall, the ``phase diagram'' of edge-modes in the \(XY\) model on a finite chain follows the structure of the correlation functions of the \(XY\) model without boundaries. (c.f. Figure 1 of Ref.~\cite{franchini2005asymptotics})

\section*{Deriving the continuum limit}

Here we derive the continuum limit of the edge-mode and the Hamiltonian in the Krylov basis assuming that both the even matrix elements $b_{2n}$, and the odd ones $b_{2n-1}$ of (8) in the main text, are separately some smooth functions of $n$ in agreement with, e.g., the model (12) in the main text. Denoting the eigenstate in the Krylov basis as $\Psi_i$, we represent the eigenvalue problem as
\begin{align}
  E \Psi_{2n-1} &= b_{2n-1} \Psi_{2n} + b_{2n-2} \Psi_{2n-2},\\
  E\Psi_{2n}&=b_{2n}\Psi_{2n+1}+b_{2n-1}\Psi_{2n-1}\, .
\end{align}
We now denote \(\psi_n = (-1)^n \Psi_{2n-1}, \phi_n = (-1)^n \Psi_{2n}\),
and rewrite
\begin{align}
  E \psi_{n} &= b_{2n-1} \phi_{n} - b_{2n-2} \phi_{n-1},\\
  E\phi_{n}&=b_{2n-1}\psi_{n}-b_{2n}\psi_{n+1},\, 
\end{align}
or introducing the operator for translation $e^{\partial_n}$
\begin{align}
    E\begin{pmatrix}\psi_{n} \\ \phi_{n}\end{pmatrix} 
    = \begin{pmatrix}0& b_{2n-1}- b_{2n-2} e^{-\partial_n}  \\ b_{2n-1}- b_{2n}e^{\partial_n} &0\end{pmatrix}
    \begin{pmatrix}\psi_{n} \\ \phi_{n}\end{pmatrix} \,.
 \label{Kodd}
\end{align}
The Krylov Hamiltonian then takes the form
\begin{align}
    H_K &= (b_{2n-1}-b_{2n}e^{\partial_n})\sigma^- +h.c.
    \approx (b_{2n-1}-b_{2n}-b_{2n}\partial_n)\sigma^- +h.c. \,,
 \label{HKcont}
\end{align}
where $\sigma^\pm=(\sigma^x\pm i\sigma^y)/2$ are Pauli matrices. In the last step we Taylor expanded $e^{\partial_n}\approx 1+\partial_n$ assuming a smooth dependence of $\psi_n,\phi_n$ on $n$. 

Let us now use the continuum version of the Krylov Hamiltonian to find an approximate zero mode $H_K\Psi=0$. We find
\begin{align}
    \Psi_n \sim \exp\left\{-\int^n dm\, \frac{b_{2m}-b_{2m-1}}{b_{2m}}\right\}\begin{pmatrix} 1 \\ 0\end{pmatrix} \,.
\end{align}
This zero mode is normalizable if the integral converges as $n\to \infty$. 
For the model (12) in main text, we have
\begin{align}
    \Psi_n \sim \exp\left\{-\int^n dm\, 
    \frac{\delta-2\alpha(1-\rho)m}{\alpha_0+\delta+2\alpha\rho m}\right\}
    \begin{pmatrix} 1 \\ 0\end{pmatrix} \,.
\end{align}
One can clearly see that the wave function $\Psi_n$ decays while $n<N^*=\frac{\delta}{2\alpha(1-\rho)}$ and then grows after that. At the minimum 
\begin{align}
    \Psi_{N^*} \sim \exp\left\{-\frac{\delta}{2\alpha}\ln\frac{1}{1-\rho}\right\},
\end{align}
reproducing the estimate (13) of the main text.

The continuum limit presented here illustrates the role of the staggering of the Krylov hopping amplitude $b_i$ for the existence of a zero mode. Indeed, if $b_{2n-1}=\alpha_0$ and $b_{2n}=\alpha_0+\delta$ we obtain $H_K=\sigma^y(\alpha_0+\delta)i\partial_x -\sigma^x \delta $. This is nothing else but the one-dimensional Dirac Hamiltonian with the mass $b_{2n-1}-b_{2n}=-\delta$. For $\delta>0$ it possesses a mid-gap state bound to the left spatial boundary. For the hopping model (12) of the main text, at very large $n$, the mass changes sign as we have $b_{2n-1}-b_{2n}\approx \alpha(1-\rho)2n>0$. If this sign change of the mass happens only at large $n$ (guaranteed by the smallness of $\alpha$ in (12) in the main text), there still exists a mode almost localized at the left end of the chain which translates to the unusually long decay of the autocorrelation function (5) in the main text. 

\section*{Derivation of Eq.~(16) in the main text}

Here we present some details on a heuristic argument justifying the estimate (16) of the main text, for the 
decay-rate. The argument is very similar to the one presented in Ref.~\onlinecite{Yao19} for a different model. 

Let us start with the Hamiltonian (1) of the main text, with $\gamma=1$ (Ising limit), 
\begin{align}
  H = J N + g \sum_i \sigma^z_i + J_z\sum_{i}\sigma^z_i \sigma^z_{i+1};\;\;\; N =\sum_i \sigma^x_i\sigma^x_{i+1}.
 \label{ham32}
\end{align}
We assume that $J\gg g\gg J_z$. The main term $N$ counts the number of domain walls in the basis of eigenstates of $\sigma^x_i$ operators. The corresponding energy for each domain wall is $-2J$. The operator $\sigma_i^z$ for $i\neq 1$ changes the number of domain walls by $0, \pm 2$ with the corresponding energy change being $0,\pm 4J$. 
The perturbation $g\sum_i\sigma_i^z$ cannot alone relax the boundary spin as flipping the boundary spin creates just one domain wall whose energy cost is $\pm 2J$, and this is off resonant by \(2J\) 
with respect to creating a bulk domain wall. This is essentially why (\ref{ham32}) with $J_z=0$ has an exact strong zero mode. Let us consider now the case of small but non-vanishing $J_z$.  

We start by setting \(g=0\), and consider only the effect of the \(J_z\) term. We would like to recast 
\(H = J N + J_z D + J_z \tilde{E}\) where \(D\) commutes with \(N\) and \(\tilde{E}\) does not.
Extending the domain wall counting argument of Ref.~\onlinecite{Nayak17}, we note that since \(N\) counts the number of domain walls, \(D\) should be such that it does not change the number of domain walls.

It is easy to see that one can take 
\begin{align}
  D = \sum_j P_j \sigma^z_j \sigma^z_{j+1}\,, \qquad \tilde{E} = \sum_j(1-P_j)\sigma^z_j \sigma^z_{j+1}\,,
\end{align}
where $P_j$ is the projector operator given by
\begin{align}
  P_j = \frac{1}{2}\biggl[1-\sigma^x_{j-1}\sigma^x_{j}\sigma^x_{j+1}\sigma^x_{j+2}\biggr].
 \label{Pj}
\end{align}
Indeed, this projector allows only for the following 4 configurations $(+,+,+,-),(+,+,-,+),(+,-,+,+),(-,+,+,+)$ and 4 others obtained by sign inversion. Acting on any of these configurations by $\sigma_j^z\sigma_{j+1}^z$ reverses the sign of middle two sites $j$ and $j+1$ and does not change the number of domain walls on the segment from $j-1$ to $j+2$. Therefore, the commutator $[D,N]=0$ as can also be checked by a direct but somewhat cumbersome computation of the commutator.

Let us now consider the full Hamiltonian
\begin{align}
  H &= J N + J_z D + g\sum_i \sigma^z_i +J_z \tilde{E}.
\end{align}
The ``unperturbed'' Hamiltonian $JN+J_zD$ preserves the number of domain walls and describes hard core domain walls moving by jumping across two sites with the probability amplitude $J_z$.\footnote{This motion of domain walls makes our model very different from the one considered in Ref.~\cite{Yao19}.} As a result the diagonalization of this Hamiltonian should lead to a dispersion of domain walls with band-width $J_z$, the minimal cost of creating the domain wall being $2J-J_z$. A typical energy of the domain wall ``particle-hole'' pair is then $\sim J_z$.

Now, the mismatch in energy $2J$ created by the flipping of the boundary spin can be compensated by creation of $\sim J/J_z$ domain wall particle-hole pairs. As $g\gg J_z$  it is much more effective to create these pairs by the perturbation $g\sum_i\sigma_i^z$ rather than by $J_z \tilde{E}$. 

Thus, the estimate for the decay-rate is given by Eq.~(16) of the main text, which is reproduced here for convenience
\begin{align}
    \Gamma \sim g \biggl[\frac{g}{J}\biggr]^{cJ/J_z},\; c=O(1).
\end{align}
This argument produces the coefficient $c$ as a number of order of 1, which is consistent with our numerical results. However, because of the heuristic character of the presented argument, we cannot rule out logarithmic corrections~\cite{Nayak17,Abanin17b}.

\end{widetext}


\end{document}